\documentclass[twocolumn,aps,prd,superscriptaddress,nofootinbib,titlepage]{revtex4-1}

\usepackage{graphicx}
\usepackage{amsmath,amsfonts}
\usepackage[hypertex]{hyperref}

\newcommand{\bea}{\begin{eqnarray}}
\newcommand{\eea}{\end{eqnarray}}
\newcommand{\ignore}[1]{}



\begin{document}

\title{Constraints on W Prime Models for the $t\bar{t}$ Asymmetry} 

\author{Christian Spethmann}

\affiliation{Physics Department, Boston University, Boston MA 02215, cspeth@bu.edu}

\begin{abstract}
\vskip 3pt \noindent
This note is a synopsis of a talk given at the SUSY 2011 conference at Fermilab on September 1, 2011.
We discuss if the Tevatron $t\bar{t}$ asymmetry can be explained by T-channel exchange of a 
$W'$ gauge boson coupling to down and top quarks. 
In a spontaneously broken gauge theory, such a $W'$ is necessarily accompanied by a $Z'$
at a similar mass scale. Null results from Tevatron searches for dijet and dilepton resonances 
imply large mass splitting between the $W'$ and $Z'$. We argue that such splitting 
can only be accomplished if the gauge symmetry is broken by a scalar in a large dimension 
representation of the gauge group, for which no perturbative description exists.
\end{abstract}

\maketitle


\section{Introduction}

Experiments at the Tevatron have observed a forward-backward 
asymmetry in $t\bar{t}$ pair production exceeding the NLO QCD prediction 
of 0.06 $\pm$ 0.01 \cite{Kuhn:1998jr, Kuhn:1998kw, Bowen:2005ap, Almeida:2008ug}. 
The CDF collaboration found that the asymmetry is 
especially pronounced in the high invariant mass bin with $M_{t\bar{t}} \ge$ 450 GeV. 
In this bin asymmetries of 0.475 $\pm$ 0.114  and 0.212 $\pm$ 0.096 
were measured in the semileptonic \cite{Aaltonen:2011kc} and dileptonic modes \cite{CDF_Note_10436}, 
respectively. The Standard Model prediction as obtained with the {\sc mcfm} code is 0.088 $\pm$ 0.013.

Taking into account the total $p\bar{p} \to t\bar{t}$ cross section and 
event reconstruction efficiencies, it was found in \cite{Gresham:2011pa} that T-channel 
exchange of flavor-changing $Z'$ and $W'$ bosons can produce a sufficient 
asymmetry while fulfilling all experimental constraints.
However, flavor-changing $Z'$ exchange implies the production
of same-sign top pairs at hadron colliders, which faces strong limits
from the Tevatron and the LHC \cite{Berger:2011ua, Cao:2011ew, Gupta:2010wx, 
Collaboration:2011dk}. 
One solution to the same sign top production problem has been proposed in \cite{Jung:2011zv}
by embedding a $Z'$ with couplings to up and top quarks in a non-Abelian flavor symmetry. 
It was found that all experimental constraints can be avoided if the $Z'$ couplings 
are nearly diagonal in the mass eigenstate basis. In this talk we address the question 
if a similar gauge theory with a $W'$ coupling to down and top quarks can explain the observed 
$t\bar{t}$ asymmetry.

\section{Gauge Symmetry Constraints on $W'$ Models}

\subsection{Mixing with Hypercharge}

Explaining the Tevatron $t\bar{t}$ asymmetry with $T$-channel exchange of a $W'$ boson
requires that down and top quarks transform into each other under the action of a 
non-Abelian symmetry group. Because of $SU(2)_L$ gauge invariance, this rotation can
not involve left-handed fields. 

The minimal realization of this symmetry is therefore $SU(2)_R$, under which the right-handed 
quark fields transform as a doublet. Since the electric charges of the down and top quark 
differ by one unit, this new symmetry can not be added to the Standard Model as an independent 
$SU(2)$ factor. The simplest symmetry breaking structure
\[ SU(2)_R \to U(1)_Y \]
produces two charged $W'$ bosons, but incorrectly predicts 
hypercharges of $\pm 1/2$ for $t^c$ and $d^c$. An additional $U(1)$ factor
\[ SU(2)_R \times U(1)_X \to U(1)_Y \]
is therefore required, implying the existence of a neutral $Z'$ boson. 
This $Z'$ has flavor-diagonal couplings and does not contribute to the 
Tevatron $t\bar{t}$ asymmetry or to same-sign top production at the Tevatron and the LHC.
Experimental limits on the $Z'$ originate from searches for dijet and $t\bar{t}$ resonances 
and the search for its suppressed dileptonic decay modes.

To reproduce the $t\bar{t}$ asymmetry, 
the $W'$ coupling constant $g_R$ is required to be large compared to the standard
model hypercharge coupling $g'$. The relation
\begin{equation}
\frac1{{g'}^2} = \frac1{g_R^2} + \frac1{g_X^2} 
\end{equation}
then implies $g_X \approx g'$ and therefore small mixing between $SU(2)_R$ and 
$U(1)_X$. If the symmetry is broken by the VEV of a scalar doublet as in the 
Standard Model, the $Z'$ is approximately degenerate with the $W'$. 

\begin{table}
\caption{\label{tab:charges} $SU(2)_R$ and $U(1)_X$ charges of Standard Model fermions in the minimal 
$W'$ model.}
\begin{ruledtabular}
\begin{tabular}{ccc}
field & $X$ & $T^3_R$ \\ \hline
$t_R$ & 1/6 & +1/2 \\
$d_R$ & 1/6 & -1/2 \\
others & Y & 0 \\
\end{tabular}
\end{ruledtabular}
\end{table}

\subsection{Couplings to Fermions}

In the limit of zero neutral gauge boson mixing, the $W'$ and $Z'$ couplings are 
\begin{align}
\mathcal{L}_{\mathrm{int}} 
& \supset  \, \begin{pmatrix} \bar{t} & \bar{d} \, \end{pmatrix} \, \gamma^\mu \left[
\, \partial_\mu - i g \frac{\sigma_a}2 {W_R^a}_\mu 
\, \right] \, P_R \,
\begin{pmatrix} t \\ d \end{pmatrix} \nonumber \\
& \supset -i g \frac1{\sqrt{2}} \, \bar{t} \gamma_\mu P_R d \, {W_R^+}_\mu 
- i g \frac1{\sqrt{2}} \, \bar{d} \gamma_\mu P_R t \, {W_R^-}_\mu \nonumber \\
& \quad - i g \frac12 \, \left( \bar{t} \gamma_\mu P_R t 
- \bar{d} \gamma_\mu P_R d \right) \, {W_R^3}_\mu
\end{align}
In this limit, the $Z'=W_R^3$ couples only to $d \bar{d}$ and $t \bar{t}$. 

As the $W'$ coupling is however finite, some degree of mixing with the 
hypercharge $U(1)_X$ is unavoidable and indeed necessary to obtain a charged gauge boson.
To calculate the branching fraction due to hypercharge admixture, 
let us first note that the gauge boson mixing matrix is 
\begin{equation} 
\begin{pmatrix} Z' \\ B \end{pmatrix}
=
\begin{pmatrix} \cos \theta_R & - \sin \theta_R \\ \sin \theta_R & \cos \theta_R \end{pmatrix}
\begin{pmatrix} W^3_R \\ X \end{pmatrix},
\end{equation}
where $X$ is the $U(1)_X$ hypercharge gauge boson and 
\begin{equation}
\sin \theta_R = \frac{g_X}{\sqrt{g_X^2+g_R^2}}, \quad \cos \theta_R = \frac{g_R}{\sqrt{g_X^2+g_R^2}}. 
\end{equation}
The fermion $Z'$ current is then
\begin{equation}
\label{eq:fermion_charges} 
J_{Z'}^\mu(f) = \sum_f  \left( g_R \cos \theta_R T^3_R 
- g_X \sin \theta_R X \right) f^\dagger \bar{\sigma}^\mu f . 
\end{equation}
with fermion charges as listed in Table \ref{tab:charges}.

\begin{figure}
\centering
\includegraphics[width=0.45\textwidth]{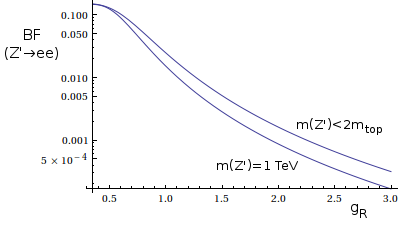}
\caption{$Z'\to e^+e^-$ branching fraction as a function of $g_R$ in the tree level approximation. 
All fermion masses except $m_{\mathrm{top}}$ are assumed to be zero.}
\label{fig:BRee}
\end{figure}

In the general case of non-zero left-handed and right-handed couplings
\begin{equation}
\mathcal{L}_{\mathrm{int}}\supset \sum_f \, i \bar{f}\gamma_\mu (k_R P_R+k_LP_L)f {Z'}^\mu ,
\end{equation}
the partial decay width of the $Z'$ into any type of fermion pairs is
\begin{align}
& \Gamma(Z'\to f\bar{f}) = \frac{N_C}{24 \pi} M_{Z'} \sqrt{1-4 \zeta^2} \nonumber \\
& \qquad \times \left[ (k_L^2+k_R^2) (1-\zeta^2) + 6 \zeta^2 k_Lk_R \right],
\end{align}
where $\zeta=m_f/M_{Z'}$.
In the interesting region of large $W'$ coupling, the $Z'$ is mostly the neutral $SU(2)_R$ gauge boson 
with only small
hypercharge admixture. The small mixing angle suppresses the branching fraction to leptons by 
\begin{equation}
\frac{\Gamma(Z' \to e^+e^-)}{\Gamma(Z' \to d \bar{d})} \approx \frac1{N_c} \, 
\frac{g_X^2 \sin^2 \theta_R}{g_R^2 \cos^2 \theta_R} = \frac13 \, \left( \frac{g_X}{g_R} \right)^4 . 
\end{equation}
Generating a sufficient asymmetry requires $g_R \gtrsim 1$, such that $g_X = g' \approx 1/3$. 
The leptonic decay modes of the $Z'$ are then suppressed by more than two orders of 
magnitude compared to the hadronic modes (see Fig.\ref{fig:BRee}).

\subsection{Embedding into Left-Right Model}

Coupling the $W'$ only to right-handed down and top quarks is sufficient to generate
a $t\bar{t}$ forward-backward asymmetry at the Tevatron. A possible objection to
this construction is the ad-hoc flavor structure. We therefore briefly note that it
is possible to embed the above theory into a left-right model with maximal mixing between 
the first and third family of right-handed quarks and asymmetric couplings to left and 
righthanded gauge bosons. The strong constraints from kaon mixing on the mass of the $W'$ boson 
\cite{Langacker:1989xa, Buras:2010pz} can be avoided by fine-tuning  
the right-handed analog of the CKM matrix to the off-diagonal unitary form
\begin{equation}
V^R = (U_u^R)^\dagger U_d^R = \begin{pmatrix} 0 & 0 & 1 \\ 0 & 1 & 0 \\ 1 & 0 & 0 \end{pmatrix}
\end{equation} 
in the mass basis, up to small additional mixing between the families.

\section{Tevatron Resonance Searches}

At hadron colliders, the $Z'$ is produced as a resonance from a $d\bar{d}$ 
initial state. The $Z'$ is therefore constrained by 
searches for dijet resonances \cite{Aaltonen:2008dn} and 
top pair production resonances \cite{Abazov:2008ny}.

Leptonic decays of the $Z'$ add additional constraints on the model. As noted above, 
the decay width to leptons is due to hypercharge admixture and 
suppressed by the small mixing angle. However, the limits on 
$p \bar{p} \to Z' \to e^+ e^-$ from Tevatron searches are in the 
femtobarn range \cite{Abazov:2010ti} and therefore severely restrict the model 
parameter space. They become especially relevant for small $Z'$ mass (i.e. 
large production cross section) and large $U(1)_X$ coupling constant, i.e. large 
mixing and therefore large branching fraction to dileptons. 

\begin{figure}
\centering
\includegraphics[width=0.48\textwidth]{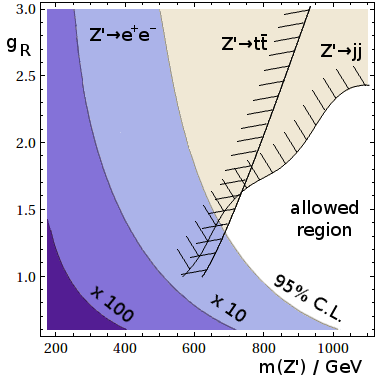}
\caption{Tevatron constraints on the simplified $SU(2)_R$ model
in $Z'$ mass vs. $SU(2)_R$ coupling parameter space. The shaded contours show the 95 \% C.L. on 
$\sigma(p\bar{p} \to Z' \to e^+e^-)$ from Tevatron dielectron resonance searches
as well as 10 and 100 times the excluded cross section $\times$ branching fraction.}
\label{fig:AllConstraints}
\end{figure} 

Assembling constraints from Tevatron $Z' \to$ dijet, $t\bar{t}$ and $e^+e^-$ searches, we find that
the dielectron search is complimentary to the search for hadronic decays at small $Z'$ mass. 
The $SU(2)_R$ model $Z'$ is therefore excluded for all couplings up to 700 GeV (Fig.\ref{fig:AllConstraints}).
At the corresponding $SU(2)_R$ coupling of $g_R = 1.4$, the $W'$ mass has to be as light as 200 GeV to 
produce a sufficient $t\bar{t}$ asymmetry. 

If the $W'$ originates from a left-right model as described above, 
all six quark flavors are charged under $SU(2)_R$. The branching fraction to 
leptons at large $g_R$ is then reduced by a factor of 1/3. However, the conclusions 
from Tevatron dilepton searches are virtually unchanged because of the additional 
$u\bar{u} \to Z'$ production channel. 

A light ${W'}$ with couplings to leptons is strongly excluded by
$W' \to \ell + X$ searches. Such a universally coupling $W_R'$ at the 
TeV scale mass is however an attractive search target for the LHC \cite{arXiv:1011.5918}.

\section{Large Higgs representations}

Explaining the Tevatron asymmetry with a $W'$ requires splitting the  
mass of the $Z'$ and $W'$ bosons by at least a factor of three. Since the mixing angle between the 
hypercharge and $SU(2)_R$ gauge groups must be small, the only way this can be accomplished 
is with a scalar VEV in a large representation of $SU(2)_R$. In the following section we will briefly
outline why such a symmetry breaking setup is disfavored by theoretical arguments.

Let us assume that the right-handed gauge symmetry is broken by a VEV in the lowest weight 
component of a scalar $\Phi_N$ transforming in the complex, dimension $N$ (i.e.~isospin $s=(N-1)/2$) 
representation of $SU(2)_R$. Neglecting the small mixing with hypercharge, the gauge boson masses are
\begin{equation}
M_{Z'} = \frac{gf}{2} \, (N-1), 
\qquad 
M_{W'} = \frac{gf}{2} \, \sqrt{N-1} .
\end{equation}
Additionally, a doublet Higgs $\Phi_2$ is required to generate masses for the top and down quarks. Ignoring this for the
moment, we can ask what size of the Higgs representation is necessary to reconcile the Tevatron $Z'\to jj$, $Z' \to t\bar{t}$ and
$Z' \to e^+e^-$ constraints with the asymmetry from T-channel $W'$ exchange.
Not overproducing $t\bar{t}$ at large invariant mass while generating a sufficient asymmetry 
requires a $W'$ with a mass $m_{W'}=200$ GeV and a coupling $g_R=1.4$. Since the $Z'$ is 
excluded by dijet constraints up to 700 GeV, we find
\begin{equation}
\sqrt{D-1} \, > \, \frac{700 \mbox{ GeV}}{200 \mbox{ GeV}} 
\quad
\Rightarrow
\quad
D \ge 14.
\end{equation}
The VEV of the doublet $\Phi_2$ contributes equally to the $W'$ and $Z'$ masses and requires the dimension of 
$\Phi_N$ to be even larger than estimated above.

\begin{figure}
\centering
\includegraphics[width=0.3\textwidth]{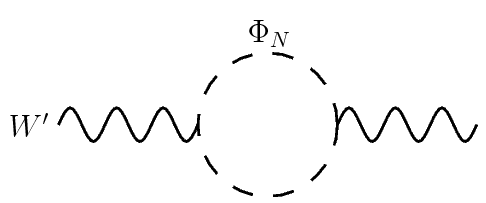}
\caption{Scalar loop contribution to the $W'$ and $Z'$ gauge boson propagators.}
\label{fig:Loop}
\end{figure} 

$\Phi_N$ is charged under the $SU(2)_R$ gauge group and therefore contributes  
to the gauge boson propagator (Fig. \ref{fig:Loop}) with strength
\begin{equation}
\sim \frac{g_R^2}{16 \pi^2} \frac{\mbox{Tr }(T^a_DT^a_D)}{\mbox{Tr } \Box} 
= \frac{g_R^2}{16 \pi^2} \frac1{12} \, (D^3-D).
\end{equation}
For $D \ge 13$ and $g_R=1$, the scalar loop contribution to the gauge boson
propagator is comparable to the tree level term, implying the breakdown
of perturbativity and the loss of predictive power of the theory.

\section{Conclusions}

By combining gauge symmetry arguments and negative Tevatron $Z'$ resonance search results, 
we conclude that the Tevatron $t\bar{t}$ asymmetry can not be explained by
T-channel exchange of a weakly coupled $W'$ gauge boson. 

Unitarity bounds on partial wave $VV \to VV$ and $Vf \to Vf$ scattering amplitudes require 
any theory of massive vector bosons to be either equivalent to a spontaneously broken gauge 
theory or result from strongly coupled new physics \cite{Cornwall:1974km}. 
The momentum scale at which new states are needed to unitarize scattering amplitudes 
can be estimated by 
\begin{equation} 
\frac{s}{4M_{W'}^2} = \left( \frac{g^2}{4\pi} \right)^{-1} .
\end{equation}
Generating a sufficiently large asymmetry requires $m_{W'}=$ 200 GeV and 
$g_R \approx 1$, implying \mbox{$\sqrt{s} \approx$ 1 TeV}. The absence of additional 
operators generated by strong dynamics can then not be explained 
except by fine tuning. \footnote{This line of argument is developed further 
in the recently published article \cite{arXiv:1111.5021}.}

\begin{acknowledgments}
We would like to thank Andy Cohen, Martin Schmaltz and Brock Tweedie for helpful conversations.
This work is supported by the Department of Energy under grant DE-FG02-01ER-40676. 
\end{acknowledgments}

\end{document}